\documentclass[9pt,twocolumn]{article}
\usepackage[left=.75in,right=.75in]{geometry}
\usepackage{booktabs}
\usepackage{dsfont}
\usepackage{amssymb}

\usepackage{multirow}
\usepackage{tikz}
\usepackage{amsmath}
\usepackage{xurl}
\usepackage{threeparttable}
\usepackage{multirow}
\usepackage{placeins}
\usepackage{subcaption}

\usepackage{xcolor}
\usepackage[linesnumbered,ruled,vlined]{algorithm2e}
\usepackage[all]{xy}
\usepackage{hyphenat}
\usepackage{tablefootnote}
\usepackage{titling}
\usepackage{balance}
\usepackage[square,numbers]{natbib}

\usepackage[utf8]{inputenc} % allow utf-8 input
\usepackage[T1]{fontenc}    % use 8-bit T1 fonts
\usepackage{hyperref}       % hyperlinks
\usepackage{url}            % simple URL typesetting
\usepackage{booktabs}       % professional-quality tables
\usepackage{amsfonts}       % blackboard math symbols
\usepackage{nicefrac}       % compact symbols for 1/2, etc.
\usepackage{microtype}      % microtypography
\usepackage{lipsum}
\usepackage{fancyhdr}       % header
\usepackage{graphicx}       % graphics
\graphicspath{{media/}}     % organize your images and other figures under media/ folder
\usepackage{xcolor}
\usepackage{threeparttable}
\usepackage{booktabs}
\usepackage{multirow}
\usepackage{amsmath}
\usepackage{hyperref}
\usepackage[capitalize]{cleveref}
\usepackage{makecell}
\usepackage{bm}
\usepackage{bbm}
\pretitle{\centering\Large\bfseries}
\posttitle{\par\vspace{1em}}

\preauthor{\centering\normalsize\lineskip 0.5em}
\postauthor{\par}

\predate{\centering\small}
\postdate{\par\vspace{1.5em}}

\begin{document}

\newcommand{\todo}[1]{\textbf{\textcolor{red}{TODO: #1}}}
\newcommand{\holdout}{D_{\text{holdout}}}
\newcommand{\aux}{D_{\text{aux}}}
\newcommand{\target}{D_{\text{target}}}
\newcommand{\synthetic}{D_{\text{synthetic}}}
\newcommand{\dataset}{D_{\text{target}}}
\newcommand{\model}{M}
\newcommand{\x}{x}
\newcommand{\shadow}{D_{\text{shadow}}}
\newcommand{\attack}{\varphi}
\newcommand{\test}{\tau}
\newcommand{\dcrtest}{\tau_{\text{DCR}}}
\newcommand{\nndrtest}{\tau_{\text{NNDR}}}
\newcommand{\imstest}{\tau_{\text{IMS}}}
\newcommand{\dcrmeasure}{\mu_{\text{DCR}}}
\newcommand{\metrictest}{\tau_{\text{DCR,NNDR,IMS}}}
\newcommand{\dcrdist}{d_{\text{DCR}}}
\newcommand{\nndrdist}{d_{\text{NNDR}}}
\newcommand{\imsdist}{d_{\text{IMS}}}

\title{The DCR Delusion: Measuring the Privacy Risk of Synthetic Data}

\date{}

\author{
    Zexi Yao\textsuperscript{1,*} \quad
    Nata\v sa Kr\v co\textsuperscript{1,*} \quad
    Georgi Ganev\textsuperscript{2} \quad
    Yves-Alexandre de Montjoye\textsuperscript{1} \\
    \vspace{0.5em}
    \textsuperscript{1}Imperial College London \\
    \textsuperscript{2}University College London and SAS
}

\maketitle

\def\thefootnote{*}\footnotetext{Equal contribution}\def\thefootnote{\arabic{footnote}}

\begin{abstract}
  Synthetic data has become an increasingly popular way to share data without revealing sensitive information. Though Membership Inference Attacks (MIAs) are widely considered the gold standard for empirically assessing the privacy of a synthetic dataset, practitioners and researchers often rely on simpler proxy metrics such as Distance to Closest Record (DCR). These metrics estimate privacy by measuring the similarity between the training data and generated synthetic data. This similarity is also compared against that between the training data and a disjoint holdout set of real records to construct a binary privacy test. If the synthetic data is not more similar to the training data than the holdout set is, it passes the test and is considered private. In this work we show that, while computationally inexpensive, DCR and other distance-based metrics fail to identify privacy leakage. Across multiple datasets and both classical models such as Baynet and CTGAN and more recent diffusion models, we show that datasets deemed private by proxy metrics are highly vulnerable to MIAs. We similarly find both the binary privacy test and the continuous measure based on these metrics to be uninformative of actual membership inference risk. We further show that these failures are consistent across different metric hyperparameter settings and record selection methods. Finally, we argue DCR and other distance-based metrics to be flawed by design and show a example of a simple leakage they miss in practice. With this work, we hope to motivate practitioners to move away from proxy metrics to MIAs as the rigorous, comprehensive standard of evaluating privacy of synthetic data, in particular to make claims of datasets being legally anonymous.

\end{abstract}

\section{Introduction}
Synthetic data is a popular tool for sharing and using sensitive data, used across
fields such as medicine~\cite{ceritli2023synthesizing, damico2023synthetic}, finance~\cite{fca_synth} and public security~\cite{pastaltzidis2022data}. Synthetic data generators (SDGs) aim to learn the underlying distribution of a dataset and generate synthetic data that preserves its statistical properties while protecting the privacy of individual records.

Membership inference attacks (MIAs) are widely accepted as the standard method to empirically assess information leakage and the privacy of synthetic data~\cite{houssiau2022tapas, meeus2023achilles, stadler2022synthetic, duan2023diffusion, annamalai2024you, kong2023efficient} and machine learning models in general~\cite{carlini2022membership, pollock2024free, zarifzadehlow}, both as a direct attack, and as an upper bound on more severe threats such as reconstruction or attribute inference attacks~\cite{salem2023sok, annamalai2024linear}. MIAs evaluate the privacy of synthetic data at an \emph{individual} level by assessing the risk of a particular record's membership in the training dataset being correctly inferred by an attacker.

While MIAs are the state-of-the-art method for evaluating privacy, they typically involve training multiple generative models, leading to high computational costs and motivating the use of simpler distance-based metrics. Distance to Closest Record (DCR) and similar metrics are commonly used as a proxy for MIAs, both in commercial products~\cite{mostly2020truly, syntho2025report} and for evaluating novel methods such as diffusion models~\cite{kotelnikov2023tabddpm, zhang2024tabsyn, pang2024clavaddpm}. These metrics assess the privacy of a synthetic dataset as a whole by measuring the similarity between synthetic and training data. Intuitively, the less similar a synthetic dataset is to its training data, the stronger the assumed privacy. The metrics can be used either to construct a binary privacy test $\dcrtest$ classifying a dataset as ``private'' or ``non-private,'' or directly as a continuous measure of privacy $\dcrmeasure$. The binary test compares the distribution of distances between synthetic and training records against the distances between a set of real holdout records and the training records, typically at a certain percentile.

\paragraph{Contribution.}
In this work, we evaluate the effectiveness of DCR and similar metrics as a proxy for membership inference attacks, and show them to be an inadequate measure of both information leakage and the privacy risk of generated synthetic data. 

First, we show proxy metrics to provide a misleading measure of privacy risk for well-known classical SDGs: IndHist~\cite{ping2017indhist}, Baynet~\cite{baynet}, and CTGAN~\cite{ctgan}. Across 9 experimental setups, we generate more than 10,000 datasets and find the majority of them to be deemed private by proxy metrics as applied in industry. Yet, instantiating MIAs against outlier records in these datasets reveals significant information leakage, with records shown to be highly vulnerable to membership inference attacks (AUC$>0.8$). Worse, we show MIAs to perform equally well against datasets deemed ``private'' and ``non-private'' by DCR, and an absence of correlation between MIA performance and $\dcrmeasure$.

Second, we show our empirical results to extend to diffusion models, a more recent popular class of synthetic data generation models. For diffusion models TabDDPM~\cite{kotelnikov2023tabddpm} and ClavaDDPM~\cite{pang2024clavaddpm}, we generate synthetic datasets considered private by $\dcrtest$, and instantiate the state-of-the-art MIAs for tabular diffusion models against them. Similarly to classical models, the MIAs reach high performance (TPR at FPR=0\% above $10\%$) despite passing the binary privacy test $\dcrtest$. $\dcrmeasure$ also shows no correlation with vulnerability to MIAs.

For computational reasons, we previously focused on outlier records which are more likely to be vulnerable to MIAs in the case of classical models. We here study for the Baynet generator and Adult~\cite{misc_adult_2} dataset the risk for every record in the target dataset and show that the risk is not limited to outliers. Though the synthetic datasets pass the binary privacy test, an MIA is able to infer the membership of 20\% of the training records better than a random guess (AUC$\geq0.6$). We also show that our findings hold across different choices of the $\dcrtest$ hyperparameter, the comparison percentile. Finally, we study a real-world example of a simple privacy leakage that the proxy metrics are by-design unable to detect.

Taken together, our results show DCR and other distance-based metrics to be poor proxies for measuring privacy risk. They detect only the most severe privacy violations, such as when synthetic data consists mostly of copies of the real data, and potentially leaving more subtle information leakage undetected. They also seem to show no meaningful correlation with actual privacy risk, making them unreliable even as general indicators of privacy. We hope this work will motivate rigorous privacy evaluation using state-of-the-art attacks defined in the literature, and encourage researchers and practitioners to move away from using distance-based metrics.

\section{Preliminaries}

In this section, we introduce relevant notation, synthetic data generators, and methods for measuring privacy of synthetic data.

\paragraph{Notation.}

We denote a record consisting of $k$ attributes with $x_i = (x_{i,1}, \ldots, x_{i,k})\sim\mathcal{D}$, where $\mathcal{D}$ is the distribution over feature space $\mathcal{F}=\mathcal{F}_1\times\ldots\times \mathcal{F}_k$. $D=\{ x_1,\ldots,x_m \}$ denotes a tabular dataset where one record $x_i$ corresponds to one row. 

\paragraph{Synthetic Data Generators (SDGs).}

Let $\dataset=\{ x_1, \ldots , x_n \}$ be a real tabular dataset. A generative model $\model=\phi(\dataset)$ trained using procedure $\phi$ estimates the underlying distribution of $\dataset$. A synthetic dataset $\synthetic\sim \model$ can then be sampled from the model. In general, we consider synthetic datasets of the same size as the training data, $|\synthetic|=|\dataset|$. We distinguish two main categories of SDGs in this paper, classical and diffusion models. Classical models are older generative models that typically learn distributions of feature values in the training dataset. Diffusion models, introduced more recently, generate data through an iterative denoising process, allowing them to capture more complex data distributions and dependencies.

\paragraph{Measuring privacy of synthetic datasets.} 

The standard approach for evaluating the privacy risk of synthetic data in the literature are membership inference attacks (MIAs). They estimate the risk of a record's membership in the training data of an SDG being correctly inferred by an attacker. However, MIAs are typically computationally expensive, leading practitioners and researchers to rely on simpler distance-based metrics such as DCR as proxies. In the following sections, we introduce these methods in more detail and evaluate their effectiveness.

\section{Privacy Evaluation Techniques for Synthetic Data}\label{sec:comparison}
In this section, we describe how proxy metrics and membership inference attacks are used in practice to empirically evaluate the privacy risk of synthetic data.

\subsection{Distance to Closest Record (DCR) and Other Distance-Based Metrics}
\label{sbpms}

Proxy privacy metrics use a notion of distance between a synthetic dataset $\synthetic \sim \model(\dataset)$ and its corresponding training dataset $\dataset$. They then use this metric to either to construct a binary privacy test or as a continuous measure of privacy.

DCR is the one of the most popular metrics used for evaluating privacy risk of synthetic data in both industry~\cite{mostly2020truly, syntho2025report, ydata2023how, syntegra2021fidelity, aws2022how} and academia~\cite{damico2023synthetic, lu2019empirical, guillaudeux2023patient, sivakumar2023generativemtd, venugopal2022privacy,yale2019assessing, yoon2023ehr,borisov2023language, kotelnikov2023tabddpm, zhang2024tabsyn,zhang2023generative, zhao2021ctab, shi2025tabdiff, pang2024clavaddpm}. It defines a vector of per-record distances between datasets $D_1$ and $D_2$, where each entry is the distance from a record in $D_1$ to its nearest neighbor in $D_2$:
\begin{equation*}
    \dcrdist(D_1,D_2) = \{ \min_{x_j\in D_2} dist(x_i, x_j) \}_{i=1}^{|D_1|}
\end{equation*}
where $dist(x_i, x_j)$ could be any distance metric but is typically the sum of euclidean distance for continuous features and hamming distances for categorical features between $x_i$ and $x_j$ \cite{mostly2020truly, aws2022how, shi2025tabdiff, kotelnikov2023tabddpm}.

Other popular proxy metrics include Nearest Neighbor Distance Ratio (NNDR) and Identical Match Share (IMS). NNDR defines the distance vector $\nndrdist(D_1, D_2)$ by computing, for each record in $D_1$, the ratio between the distance to its nearest neighbor and the distance to its second-nearest neighbor in \(D_2\). Instead of a distance vector, IMS defines a scalar distance measure $\imsdist(D_1,D_2)$ as the number of records in $D_1$ with an identical match in $D_2$.

\paragraph{Privacy test.} DCR, NNDR and IMS are often used to construct binary privacy tests to classify a synthetic dataset $\synthetic$ as ``private'' or ``non-private''~\cite{mostly2020truly,aws2022how,guillaudeux2023patient}. This is done by comparing the distance between $\synthetic$ and $\dataset$ to the distance between $\dataset$ and a holdout set of real records $\holdout$. If $\synthetic$ is further away from $\dataset$ than $\holdout$, it is considered private.

DCR and NNDR construct privacy tests $\dcrtest$ and $\nndrtest$ by comparing the 5th percentile of the respective distance vectors:

\begin{align*}
\dcrtest(\synthetic,& \dataset) = \\
&\mathbbm{1}\Big[
\dcrdist(\synthetic, \dataset)_{p=0.05} \geq \\
&\dcrdist(\holdout, \dataset)_{p=0.05}
\Big]
\end{align*}

where $p=0.05$ denotes the 5th percentile. The test is defined analogously for $\nndrdist$.

IMS constructs the privacy test by comparing the number of identical matches:

\begin{align*}
    \imstest(\synthetic, \dataset) = \mathbbm{1}[ &\imsdist(\synthetic, \dataset) \leq \\&\imsdist(\holdout, \dataset) ]
\end{align*}

Here, the synthetic dataset is considered private if it contains fewer identical matches with the training data than the holdout set does.

Various combinations of these metrics are used in practice, with no agreed-upon, widely used setup. We therefore define a strict privacy test $\metrictest$ as a joint privacy test using all three proxy metric. $\synthetic$ is considered by $\metrictest$ to be privacy only if it passes all of them $\dcrtest$, $\nndrtest$, and $\imstest$.
\begin{align*}
    \metrictest(&\synthetic,\dataset) = \\ 
        &\mathbbm{1}[ \dcrtest(\synthetic, \dataset)=1\\
        &\land \nndrtest(\synthetic, \dataset)=1\\ 
        &\land \imstest(\synthetic, \dataset)=1 ]
\end{align*}
 
\paragraph{Continuous privacy measure.} DCR is also commonly used as a continuous privacy measure $\dcrmeasure$ for comparing the privacy of different generative models, particularly for diffusion models~\cite{shi2025tabdiff, kotelnikov2023tabddpm}. Instead of comparing synthetic data to a holdout set, the continuous measure aggregates the distance vector $\dcrdist(\synthetic, \dataset)$ to produce a single privacy score.
In this work, we follow~\citet{kotelnikov2023tabddpm} and use the mean of the distances in \(\dcrdist(\synthetic,\allowbreak\ \dataset)\) as the continuous privacy measure:

\begin{multline*}
    \dcrmeasure(\synthetic, \dataset) = \\ \frac{1}{|\synthetic|} \sum_{i=1}^{|\synthetic|}\dcrdist(\synthetic, \dataset)_i
\end{multline*}

A higher value of $\dcrmeasure(\synthetic, \dataset)$ indicates that synthetic records are more distant from $\dataset$ and is thus assumed to imply better privacy protection.

\subsection{Membership Inference Attacks (MIAs)}
\label{membershipinference}
MIAs are the state of the art technique for evaluating privacy risk of synthetic data~\cite{houssiau2022tapas,wu2025winning}. They identify privacy leakage using a privacy game where an attacker aims to infer whether a synthetic dataset was generated by a model trained on a specific target record. 

\paragraph{Classical models.}
The state-of-the-art MIA for classical SDGs is extended-TAPAS, a black-box attack introduced by~\citet{houssiau2022tapas} and extended by~\citet{meeus2023achilles}. Extended-TAPAS models how the inclusion or exclusion of a single target record $\x$ impacts the generated synthetic data and trains a meta-classifier to predict membership.

For a target record $\x\in\dataset$, the attacker trains shadow models on datasets sampled from an auxiliary dataset $\aux$, which is drawn from the same distribution as the target dataset $\dataset$ but is disjoint from it. The target record is included in exactly half of the shadow datasets. They then generate a synthetic dataset using each shadow model, and extract query features from each. These features count the number of synthetic records that match the target record across random subsets of attributes. This results in a labeled membership dataset for training a meta-classifier that predicts whether a given synthetic dataset was trained on the target record. In our experiments, we use $1000$ shadow models.

The MIA is evaluated across a set of evaluation models, where exactly half are trained on $\x$. In this work, we evaluate in the model-seeded setup of~\citet{guepin2024lost}, where half of the evaluation models are trained on $\dataset$, and half on $\dataset$ where $\x$ is replaced by a randomly sampled holdout record. The MIA is performed against each dataset, and its ROC AUC score is computed, resulting in a risk estimate of $\x$ within $\dataset$. In our experiments, we use $1000$ evaluation models.

As extended-TAPAS must be developed and evaluated separately for each target record, evaluating the risk of every record in $\dataset$ across setups is computationally infeasible. Because of this, in our main experiment, we select $100$ target records in $\dataset$ and instantiate the MIAs against them. We use the Achilles vulnerability score introduced by~\citet{meeus2023achilles}, and select the $100$ records with the highest vulnerability score. The final output for each setup is then a set of per-record MIA AUC scores.

\paragraph{Diffusion models.}
The state-of-the-art MIA for tabular diffusion models was introduced by~\citet{wu2025winning} in the challenge on Membership Inference over Diffusion-models-based Synthetic Tabular data (MIDST)~\cite{midst}. We consider both the black-box and white-box variant of the attack, and refer to them collectively as the MIDST attacks for simplicity. The black-box attack relies only on data generated by the target model, while the white-box attack has full access to the model and its internal parameters. Both attacks model the model’s loss on member versus non-member records.

To train the meta-classifier, the attacker first samples shadow datasets from an auxiliary dataset and trains a shadow diffusion model on each. For each shadow model, they extract features from the initial noise and training loss for both member and non-member records. These features form a labeled dataset used to train a multi-layer perceptron (MLP) classifier that predicts whether a given record was part of the training data. In our experiments, we use 20 shadow datasets to train the MIDST attacks.

As these attacks can be applied to any individual record, and training diffusion models is computationally expensive, evaluation here is typically done on a fixed target model across a set of known members and non-members. The attack is applied to each record, and a single True Positive Rate (TPR) at a fixed low False Positive Rate (FPR) is computed based on the predictions~\cite{wu2025winning, carlini2022membership}. In our setup, we evaluate the MIDST attack on 10 diffusion models, each with 200 member and 200 non-member records, resulting in 10 TPR values--one per synthetic dataset.

\section{Experimental Setup}
In this section, we specify the datasets and models we use in our experiments.

\subsection{Models}\label{genmodels}

\paragraph{Classical models.} We use three well-known synthetic data generators, using the implementations available in the reprosyn~\cite{reprosyn2022} repository. 

IndHist ~\cite{ping2017indhist} is the simplest of our selected models. It uses marginal frequency counts to generate feature values for synthetic records. For each feature, it samples from the distribution of values for that feature among all training records. Different features are sampled independently from each other.

BayNet~\cite{baynet} trains a Bayesian network to learn the relationships between features. Each feature is represented as a node on a network graph, with edges representing relations between two features. The GreedyBayes algorithm introduced by \citet{baynet} is then used to estimate the joint probabilities of the features, from which synthetic records can be sampled.

CTGAN~\cite{ctgan} trains a generative adversarial network (GAN) consisting of a generator and a discriminator to model feature distribution of records in the training dataset. They are trained jointly with opposing goals: the discriminator attempts to distinguish between real and synthetic records produced by the generator, while the generator aims to produce synthetic data similar enough to real data to fool the discriminator. 

\paragraph{Diffusion models.} Diffusion models have become increasingly popular in recent years due to their increased utility of generated synthetic data and versatility of applications compared to classical models. We use two tabular diffusion models, with the implementation available in the MIDSTModels repository~\cite{midst}.

TabDDPM~\cite{kotelnikov2023tabddpm} is the first diffusion model specifically developed for tabular data. It adapts the diffusion process to account for different feature types by applying Gaussian diffusion to numerical features and multinomial diffusion to categorical and binary features.

ClavaDDPM~\cite{pang2024clavaddpm} is a tabular diffusion model designed to generate multi-relational data. It uses latent clustering to model the relationships between the tables defined by foreign keys and enable conditional generation of synthetic tables. 

\subsection{Datasets} \label{datasets}
We evaluate the success of privacy measures across the following publicly available datasets, commonly used in literature studying tabular data privacy.

Adult~\cite{misc_adult_2} is an anonymized sample of the 1994 US Census data containing 48,842 records. It contains 15 demographic features, 9 of which are categorical.

Bank~\cite{bank_marketing_222} contains 45,211 records concerning the marketing campaign of a Portuguese banking institution in 2014. Each record contains 17 features of which 4 are demographic and 13 describe the individual's previous interactions with the institution.

UK Census~\cite{census2011} is an anonymized 1\% sample of the 2011 Census from Wales and England, published by the UK Office for National Statistics. The dataset is comprised of 569,741 records with 17 categorical demographic features.

Berka~\cite{berka} is an anonymized database containing information regarding over $5,000$ clients collected in 2000 from a Czech bank. The main dataset, which we refer to as the Berka dataset, contains over $1,000,000$ transactions. Additional tables with account and client information can be linked via foreign keys, e.g. when training ClavaDDPM.

\section{Results}
\subsection{Evaluating DCR and Other Proxy Metrics for Classical Models} \label{sec:classical}

\begin{figure*}[t]
    \centering
    \includegraphics[width=0.8\linewidth]{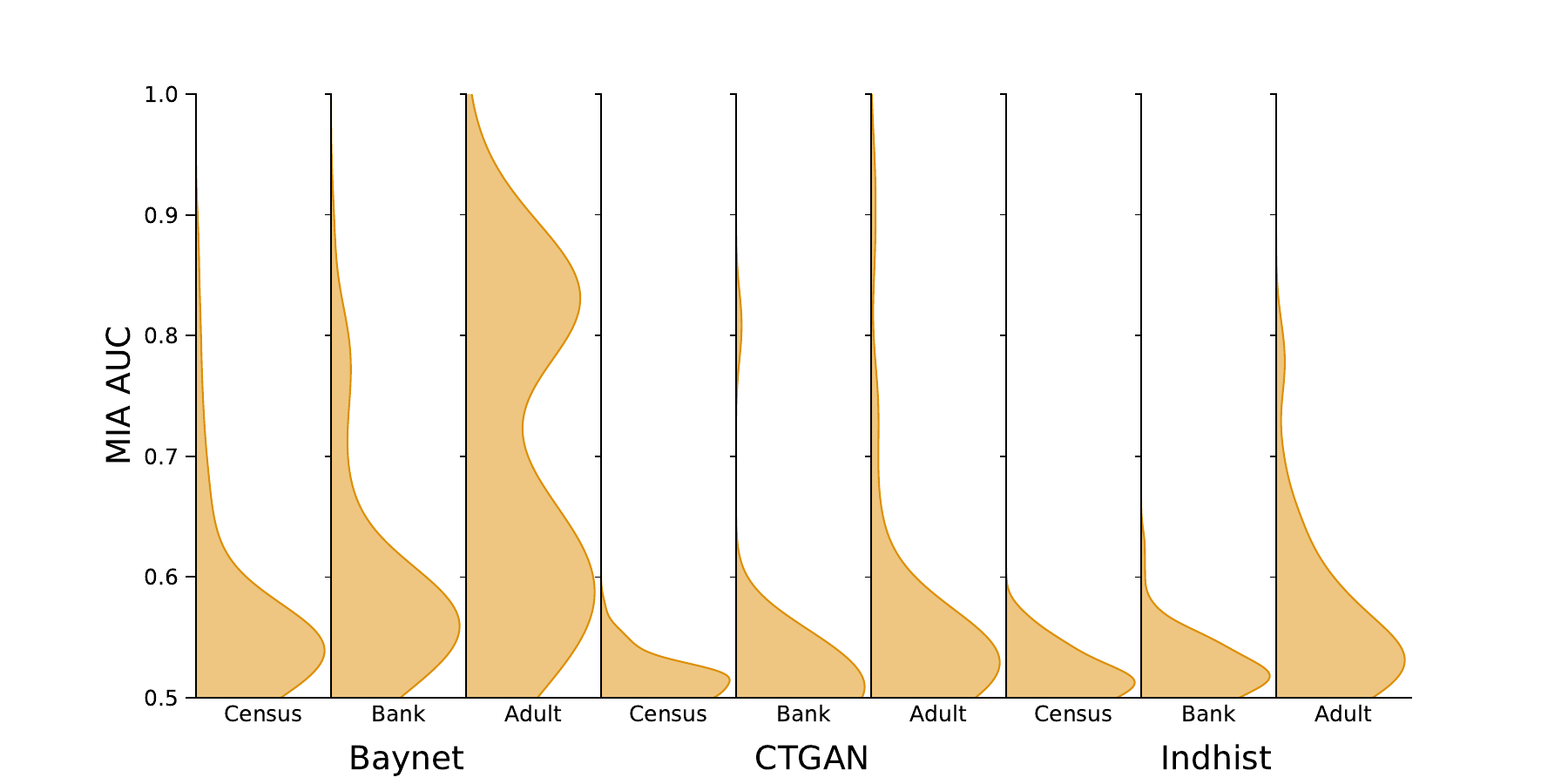}
    \caption{Extended-TAPAS MIA AUC on datasets considered ``private'' by $\metrictest$ across classical SDG setups. Each dataset-SDG setup contains 100 target records selected using the \textit{Achilles} score.}
    \label{fig:figure1}
\end{figure*}

We here evaluate the effectiveness of DCR and other proxy metrics for identifying privacy leakage in classical synthetic data generators by comparing them to MIA results across $3$ datasets and $3$ target models. We develop MIAs against $100$ outlier target records per setup, selected using the Achilles score. For each target record, we compute the percentage of evaluation synthetic records that fail $\dcrtest$ and $\metrictest$, and the mean $\dcrmeasure$ across all synthetic datasets for that record. We then study the MIA AUC values for the outlier records in datasets deemed ``private'' by $\dcrtest$ and $\metrictest$.

In $7$ out of the $9$ setups, both $\metrictest$ and $\dcrtest$ consistently classify all $500$ synthetic datasets per target record as ``private.'' The only exceptions are observed with the Baynet generator on the Census and Bank datasets. For Census, 12\% of the synthetic datasets fail the $\metrictest$ and 1.1\% fail $\dcrtest$ alone. For Bank, 0.4\% of the datasets fail the $\metrictest$, and 0\% fail $\dcrtest$. \cref{fig:figure1} shows the datasets to be highly vulnerable to MIAs, despite being considered ``private'' by the proxy metrics. 

\cref{fig:figure1} shows the datasets passing the proxy metric privacy tests to leak information about their training data. In the majority of setups, the MIA reaches $AUC\geq0.6$---shown to indicate information leakage---for a significant fraction of records, and even $AUC\geq0.8$ for some. This suggests that $\dcrtest$ and $\metrictest$ often misrepresent synthetic datasets with significant privacy leakage as ``private,'' making them unreliable for verifying privacy of synthetic data for release.

\begin{figure}
    \centering
    \includegraphics[width=0.8\linewidth]{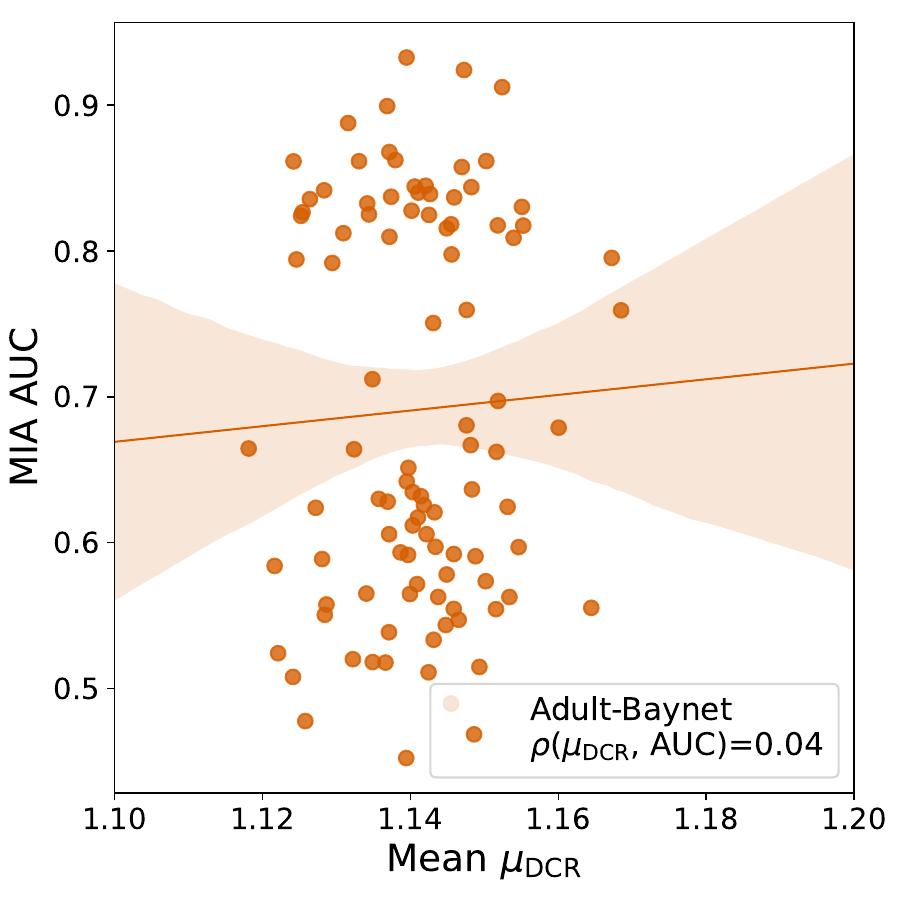}
    \caption{Comparison of mean $\dcrmeasure$ and MIA AUC for the Baynet generator on the Adult dataset. Each point represents a target record's MIA AUC and its mean $\dcrmeasure$ across evaluation datasets.}
    \label{fig:figure2}
\end{figure}

As proxy metrics often fail to flag privacy leakage, we now study whether they can still provide informative signal about the risk of a dataset. We examine whether synthetic datasets considered ``non-private'' exhibit higher MIA AUC values then those considered ``non-private.'' Then, we evaluate whether the continuous measure $\dcrmeasure$ gives an indication of MIA performance.

For the Census dataset and the Baynet generator, 12\% of the synthetic datasets are classified as ``non-private'' by $\metrictest$. We compare MIA performance when instantiated across the full set of synthetic datasets and only the ``private'' datasets. \cref{fig:figure1} shows that there is no meaningful difference in performance between the two sets -- MIA AUC remains equally high, regardless of whether evaluation is restricted to the “private” subset or not.

For each target record in the Adult-Baynet setup, we compute the corresponding MIA AUC and the mean $\dcrmeasure$ across the synthetic datasets used for evaluation and study their relationship. \cref{fig:figure2} shows that there is no correlation between the two values. AUC values span a wide range (roughly between $0.5$ and $1.0$), regardless of the value of $\dcrmeasure$, suggesting that $\dcrmeasure$ is unable to effectively distinguish between datasets with different levels of privacy risk.

\subsection{Evaluating DCR for Diffusion Models}

% \begin{table}[t]
%     \centering
%     \caption{Mean TPR@FPR=0\% across 10 target datasets per setup.}
%     \begin{tabular}{ccc}
%     \toprule
%         Dataset & Black-box attack & White-box attack \\\midrule
%         TabDDPM &$ 0.70\pm 0.03$ & $0.29\pm0.05$ \\
%         ClavaDDPM & $0.08\pm0.03$ & $0.20\pm0.08$ \\\bottomrule
%     \end{tabular}
%     \label{tab:diffusion_tprs}
% \end{table}

\begin{table}[t]
    \centering
    \caption{Pearson correlation between TPR@FPR=0\% and $\dcrmeasure$.}
    \begin{tabular}{ccc}
    \toprule
        Dataset & Black-box attack & White-box attack \\\midrule
        TabDDPM &$ 0.47$ & $0.11$ \\
        ClavaDDPM & $0.10$ & $-0.03$ \\\bottomrule
    \end{tabular}
    \label{tab:diffusion_tprs}
\end{table}

\begin{figure*}[t!]
    \centering
    \includegraphics[width=\linewidth]{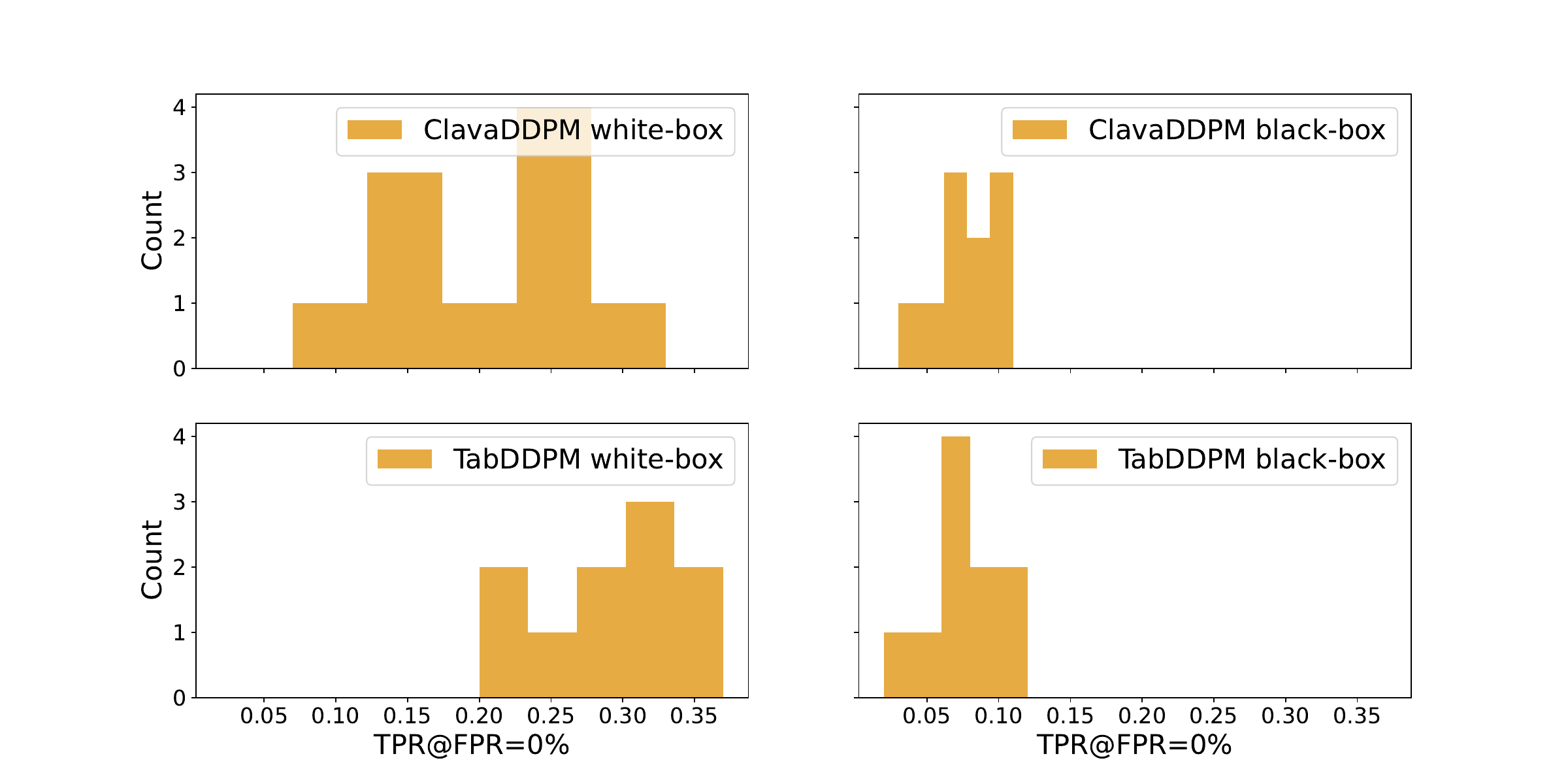}
    \caption{Distribution of TPR@FPR=0\% for MIDST attacks against TabDDPM and ClavaDDPM.}
    \label{fig:diffusion_result}
\end{figure*}

% \begin{figure}[t!]
%     \centering
%     \includegraphics[width=0.8\linewidth]{figures/diffusion/tpr_vs_dcr_clava_whitebox.pdf}
%     \caption{$\dcrmeasure$ and MIA TPR@FPR=0\% for MIDST white-box attack on $10$ ClavaDDPM target datasets.}
%     \label{fig:diffusion_dcr_measure}
% \end{figure}

We here study the effectiveness of DCR for evaluating the privacy of diffusion models by repeating the analyses done in~\cref{sec:classical}, and follow the state-of-the-art methods for diffusion models. Specifically, we focus on DCR and measure MIA performance on datasets deemed ``private'' as TPR at FPR=0\%~\cite{carlini2022membership}. Using TabDDPM and ClavaDDPM, we train $10$ target models for each method and generate one synthetic dataset per model. All generated datasets pass the DCR privacy test $\dcrtest$. We then instantiate both the black-box and white-box MIDST attacks against each of the generated synthetic datasets. 

~\cref{tab:diffusion_tprs} shows that for both TabDDPM and ClavaDDPM, the MIDST attacks are able to successfully infer membership in the training data of the target models. The black-box attack achieves TPR@FPR=0\% above 5\% on the majority of datasets, and exceeds 10\% on some datasets for both models. The white-box attack performs even better, reaching TPR@FPR=0\% above 20\% on all $10$ target datasets, and on more than half of the ClavaDDPM datasets. These results indicate clear information leakage that is not detected by $\dcrtest$.

In line with our analysis in~\cref{sec:classical}, we evaluate whether $\dcrmeasure$ provides any meaningful signal of privacy risk. We compute the $\dcrmeasure$ for all $10$ target datasets in each setup, and compare it to the TPR@FPR=0\% achieved by the MIAs. \cref{tab:diffusion_tprs} shows there to be no clear correlation between $\dcrmeasure$ and MIA performance, indicating that $\dcrmeasure$ is not a reliable proxy for privacy risk as identified by MIAs.

\subsection{Effect of Adjusting DCR Hyperparameter}

\begin{figure}[h!]
    \centering
    \includegraphics[width=0.8\linewidth]{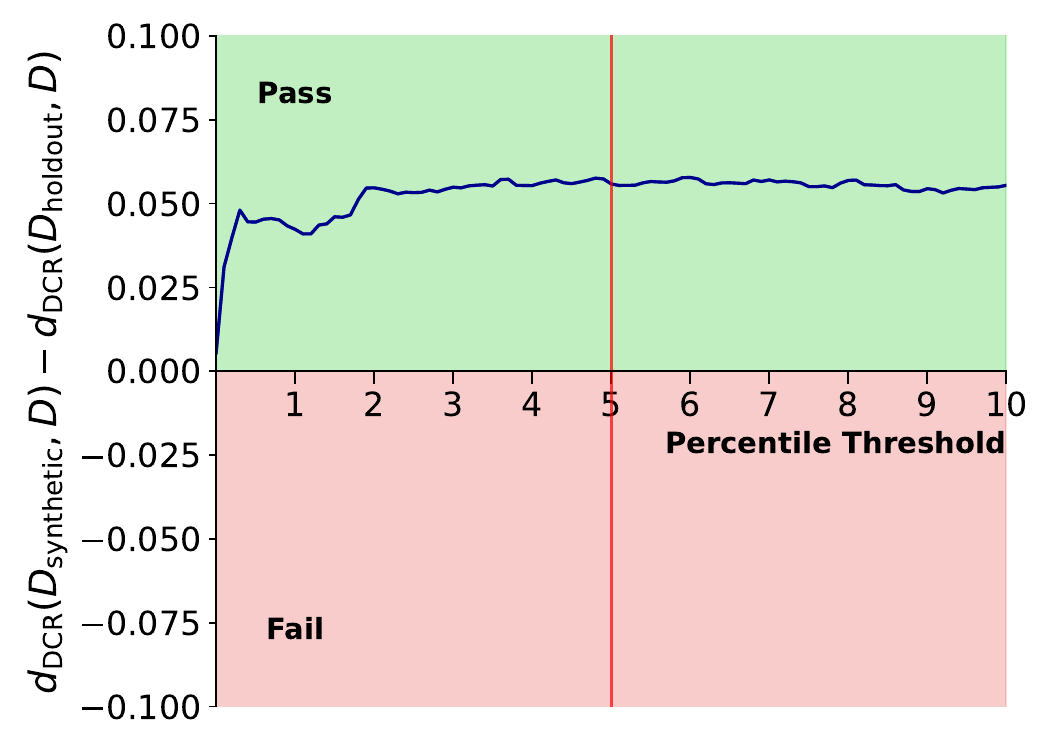}
    \caption{Comparison of $\dcrdist(\synthetic,\dataset)-\dcrdist(\holdout,\dataset)$ across percentile thresholds for a synthetic dataset trained on a vulnerable record with MIA AUC = 0.84.}
    \label{fig:dcr_thresholds}
\end{figure}

$\dcrtest$ determines the privacy of a synthetic dataset $\synthetic$ by comparing the distance vector between $\synthetic$ and $\dataset$ to the distance vector between $\holdout$ and $\dataset$ at the same percentile mark \textit{p}, typically 5th percentile. This percentile choice is the only hyperparameter of $\dcrtest$. We now study whether tuning this threshold can improve $\dcrtest$'s ability to detect privacy leakage. The condition for passing the privacy test,
\begin{equation*}
    \dcrdist(\synthetic, \dataset)_{p} \geq \dcrdist(\holdout, \dataset)_{p}
\end{equation*}
can be rewritten as:
\begin{equation*}
    \dcrdist(\synthetic, \dataset)_{p} - \dcrdist(\holdout, \dataset)_{p} \geq 0
\end{equation*}
We examine the effects of adjusting $p\in[0, 0.1]$ for the above condition across all synthetic datasets in the Baynet generator with Adult dataset setup. 

\cref{fig:dcr_thresholds} shows an example of adjusting $p$ for a single synthetic dataset trained on a vulnerable record with MIA AUC$ = 0.84$. For this synthetic dataset, the value remains above $0$ regardless of the value of $p$, showing that the dataset passes $\dcrtest$ on all thresholds $p\in[0, 0.1]$. This result holds across all synthetic datasets in the Baynet-Adult setup -- every dataset is deemed ``private’' by $\dcrtest$,  regardless of the choice of threshold.

\begin{figure}[h]
    \centering
    \includegraphics[width=0.8\linewidth]{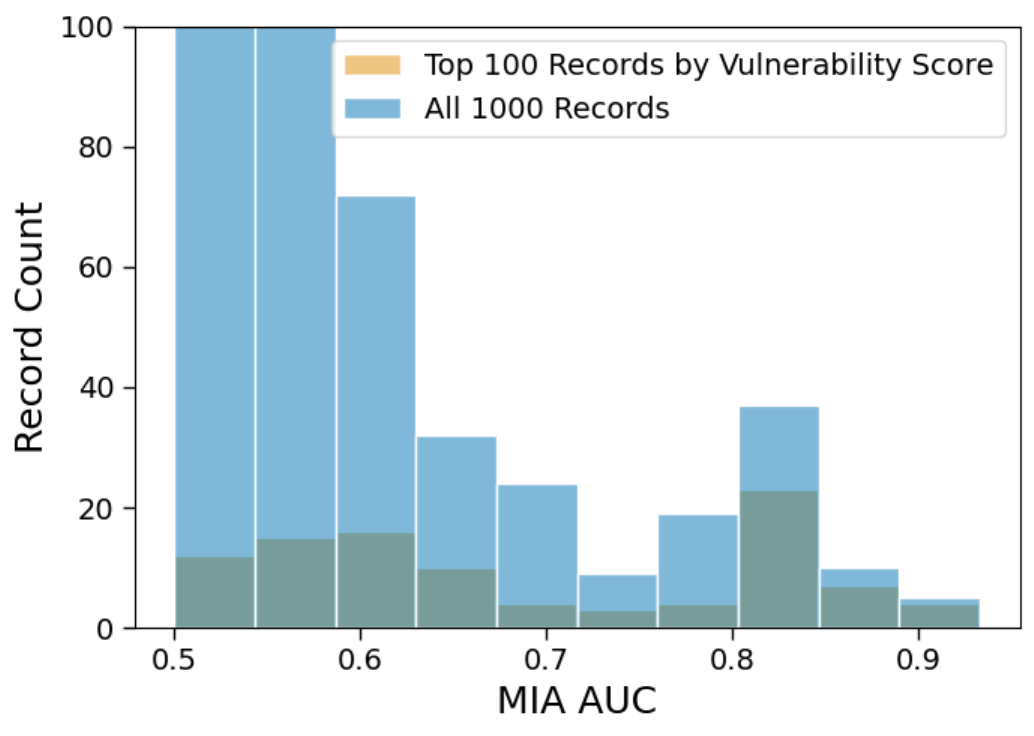}
    \caption{MIA AUCs of all 1000 target records and 100 vulnerable records selected using \textit{Achilles} in the Adult-Baynet setup, in synthetic datasets considered ``private'' by $\metrictest$.}
    \label{fig:fig4}
\end{figure}

\subsection{Analysis of the Impact of the \textit{Achilles} Vulnerability Score}
In our main experiment for classical models in \cref{sec:classical}, we reduce computational costs by only performing MIAs against the top 100 records by \textit{Achilles} vulnerability score for each dataset. To eliminate concerns that such sampling may have exaggerated the privacy risk indicated by MIA, we analyze DCR and MIA performance on all $1000$ records in $\dataset$ for one setup (Baynet with Adult). Consistent with our prior results, all synthetic datasets for all $1000$ records also pass $\dcrtest$ and $\metrictest$.

We now compare the distribution of MIA AUC values across all 1000 records to the MIA AUC values of 100 outlier records selected by the \textit{Achilles} vulnerability score in~\cref{sec:classical}. \cref{fig:fig4} shows that while \textit{Achilles} score is more likely to identify vulnerable records than random sampling, a significant proportion of vulnerable records went undetected -- the MIA achieves an AUC $\geq$ 0.8 for $54$ records and AUC $\geq$ 0.6 for $200$ records out of $1000$ total records. This is still a high percentage of records with information leakage, which indicates significant privacy risk across all synthetic datasets in the setup.

\subsection{Detailed Analysis of one Highly Vulnerable Record}

We select the record with the highest MIA AUC across all our classical setups for detailed analysis as to why DCR is unable to detect clear privacy violations. This record has an MIA AUC of $0.94$ and is from the CTGAN-Adult setup, all synthetic datasets generated by this setup passed both $\dcrtest$ and $\metrictest$. We start by identifying the cause of high privacy leakage for this record -- a distribution shift of generated synthetic datasets between CTGAN models trained on the target record and those that were not.

\begin{figure}
    \centering
    \includegraphics[width=0.8\linewidth]{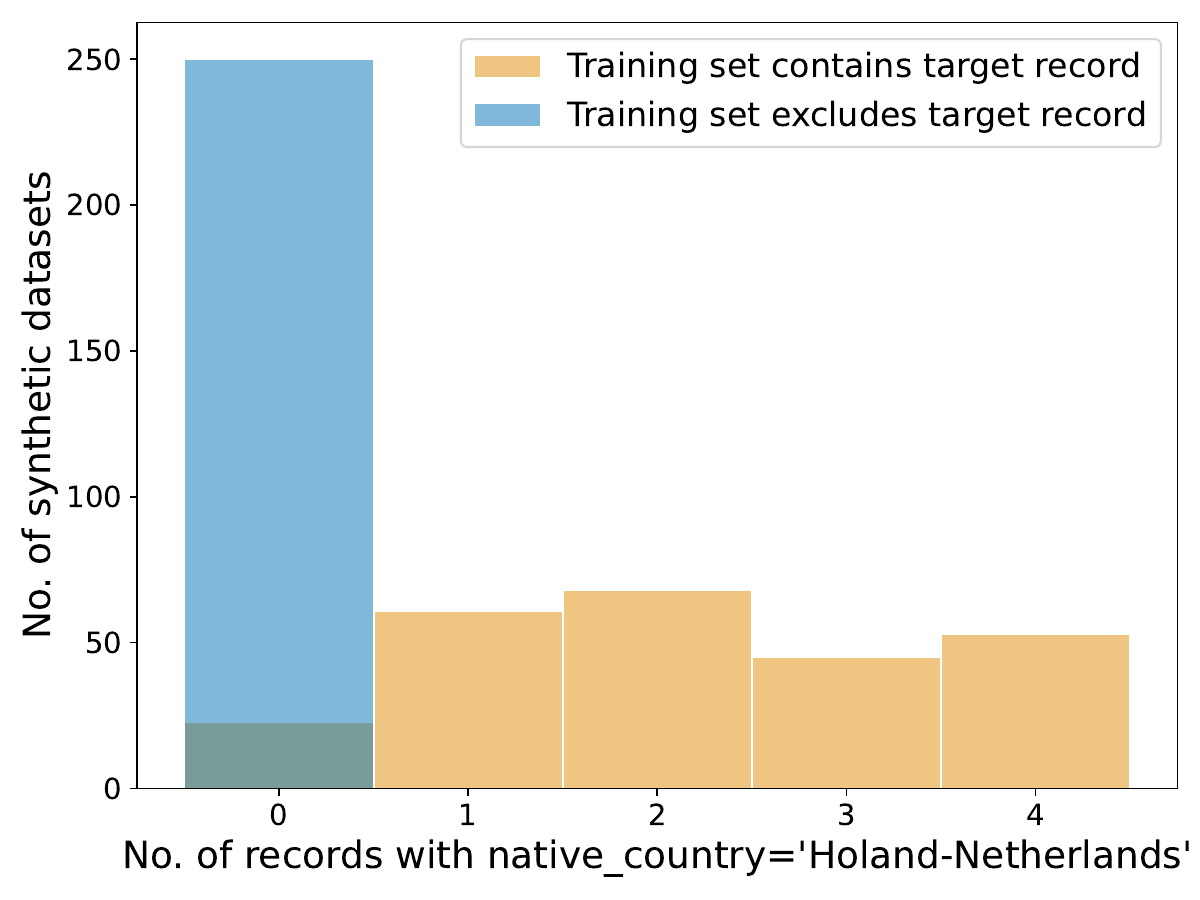}
    \caption{Comparison of synthetic datasets generated by CTGAN-Adult models trained on target record with uniquely identifying native country ``Holland-Netherlands.''}  
    \label{fig:figure3}
\end{figure}

Notably, as shown in~\cref{fig:figure3}, CTGAN models trained on this record generate synthetic data containing records with a \texttt{native-country} value of ``Holland-Netherlands'' in 92\% of cases, while models not trained on it never produce it. We immediately notice the target record is the only record in the entire Adult dataset with the value ``Holland-Netherlands'' for the \texttt{native-country} feature. Thus, the presence of a synthetic record with this feature value would reveal the membership of the target record in $\dataset$. While this is a clear privacy concern, DCR instead focuses on distance measurements to the closest record, which is not the cause of the privacy leakage. Furthermore, distance calculations treat all features with the same importance, a single feature has minimal effect on the overall distance metric -- synthetic records containing the uniquely identifying value are not even the closest records in $\synthetic$ to the target record. As a result, $\dcrtest$ and $\metrictest$ fail to flag this obvious privacy risk.

We believe this to be a core limitation of proxy metrics which by design do not learn and need to make assumptions about what causes privacy leakage. While one could design another proxy metric to check for uniquely identifying feature values, it is just a single example of privacy leakages that distance metrics cannot capture. Prior work has shown that leakage may also arise from more complex feature combinations or dataset-specific characteristics~\cite{pyrgelis2017knockknockwhosthere}, which cannot be identified by DCR and requires designing setup-specific proxy metrics. In contrast, MIAs are able to learn and thus capture a more comprehensive spectrum of privacy risks, including those that were previously unknown and unique to specific setups.

\section{Related Work}
\paragraph{Synthetic Data Generators (SDGs).} 
Numerous synthetic data generators have been proposed for tabular data~\cite{jordon2022syntheticdatawhat, cristofaro2024synthetic, hu2024sok}, spanning approaches from traditional statistical methods like graphical models~\cite{baynet, mckenna2021winning} and workload/query-based~\cite{liu2021iterative, mckenna2022aim}, to advanced deep learning techniques, including Variational Autoencoders (VAEs)~\citep{acs2018differentially, abay2019privacy} and Generative Adversarial Networks (GANs)~\cite{xie2018differentially, zhang2018differentially, jordon2018pate, ctgan}; and more recently, diffusion models~\cite{kotelnikov2023tabddpm, zhang2024tabsyn, shi2025tabdiff, mueller2025cdtd}.
As discussed in Section~\ref{genmodels}, we use a range of best-performing models with reliable and public implementations.

\paragraph{Membership Inference Attacks (MIAs).}
MIAs were initially introduced as a method to infer the presence of trace amount of DNA from released genomic aggregates~\cite{homer2008mia}. It was then extended to assess privacy leakage in discriminative machine learning models~\cite{shokri2017membership, carlini2022membership, ye2022enhanced, zarifzadehlow}.
In the context of generative models on images, \citet{hayes2019logan} and \citet{hilprecht2019monte} propose the first MIAs targeting VAEs and GANs, employing a shadow discriminator and Monte Carlo estimation, respectively.
Subsequently, \citet{chen2020gan} introduce a taxonomy of MIAs along with a model-agnostic attack against GANs.
More recently, \citet{zhu2023data} and \citet{carlini2023extracting} extend MIAs research to diffusion models, demonstrating that these models are more susceptible to memorization than GANs.

For tabular data, 
\citet{stadler2022synthetic} present the first systematic evaluation of MIAs, showing that outliers are particularly vulnerable.
Other MIAs against tabular data include TAPAS~\cite{houssiau2022tapas}, which relies on running a collection of random queries, and DOMIAS~\cite{van2023membership}, which detects overfitting using a density-based approach.
\citet{meeus2023achilles} propose an identification procedure for selecting the most vulnerable records based on distance metrics and an extension of TAPAS~\cite{houssiau2022tapas}, called extended-TAPAS.
\citet{guepin2023synthetic} relax a common assumption in MIAs that the adversary has access to an auxiliary dataset.
More recently, researchers have proposed model-specific attacks targeting traditional generative models to audit their privacy properties~\cite{annamalai2024you, ganev2025elusive, golob2025privacy}, and diffusion models to mitigate computational overhead~\cite{wu2025winning}.

\paragraph{Distance to Closest Record (DCR).}
DCR is widely adopted to measure and claim privacy in both industry~\cite{mostly2020truly, syntho2025report, ydata2023how, syntegra2021fidelity, aws2022how} and academia, particularly in the medical domain~\cite{damico2023synthetic, lu2019empirical, guillaudeux2023patient, sivakumar2023generativemtd, venugopal2022privacy, yale2019assessing, yoon2023ehr, kaabachi2025scoping}.
Furthermore, a growing number of recently proposed diffusion models -- published in top-tier ML and NLP venues such as NeurIPS, ICML, and ICLR -- rely exclusively on DCR to support privacy claims, or to demonstrate improvements over prior models~\cite{borisov2023language, kotelnikov2023tabddpm, zhang2024tabsyn, zhang2023generative, zhao2021ctab, shi2025tabdiff, pang2024clavaddpm}.

DCR is often used in conjunction with other proxy metrics like Nearest Neighbor Distance Ratio (NNDR) and Identical Match Share (IMS) in real-world synthetic data products to run statistical tests and support privacy claims~\cite{mostly2020truly, syntho2025report, aws2022how}.

This is despite existing research showing that MIAs, such as TAPAS, are more effective at detecting privacy leakage than DCR in traditional models~\cite{houssiau2022tapas, annamalai2024you}.
Additionally, \citet{ganev2025inadequacysimilaritybasedprivacymetrics} show that relying on DCR to guarantee synthetic data privacy could be dangerous as adversaries operating under strong assumptions -- such as repeated black-box access to conditional generation and proxy metric APIs -- can successfully perform MIAs and reconstruct entire training records.

\section{Discussion \& Conclusion}
Distance to Closest Record and other proxy privacy metrics are presented both as a statistical test for verifying privacy of synthetic datasets prior to data release and also as a proxy measurement of privacy of synthetic datasets~\cite{lu2019empirical, aws2022how, park2018data, yale2019assessing, zhao2021ctab, venugopal2022privacy, borisov2023language, kotelnikov2023tabddpm, liu2023tabular, yoon2023ehr, ceritli2023synthesizing, zhang2023generative, zhang2024tabsyn, panfilo2022generating, panfilo2023a, guillaudeux2023patient, sivakumar2023generativemtd, damico2023synthetic, shi2025tabdiff,kotelnikov2023tabddpm, mostly2020truly}. 

In this paper, we show across both classical and diffusion models, that DCR and other metric tests consistently fail to identify privacy leakage, including clear privacy violations such as the presence of uniquely identifying feature values. Furthermore, we also show that DCR as a proxy measurement is uninformative for comparing privacy of synthetic datasets for both classical and diffusion models -- there is no clear relation between distance of synthetic records to training dataset and MIA vulnerability.

Additionally, we show that privacy violations that are caused by a subset of feature values, such as the case of uniquely identifying feature values in the CTGAN generator with Adult dataset setup, have synthetic records that are distant from the training record. The effect of these important features on synthetic record distance is heavily reduced by the presence of other features, thus making it highly unlikely for DCR to detect such violations.

With DCR and other proxy metrics shown to be unsuitable for use as a privacy test or proxy privacy measurement, it is imperative for both the academic and industry community to move to Membership Inference Attacks, which is the state-of-the-art for measuring privacy risks of synthetic datasets.

\balance
\bibliography{mybibliography}

\begin{thebibliography}{75}
\providecommand{\natexlab}[1]{#1}
\providecommand{\url}[1]{\texttt{#1}}
\expandafter\ifx\csname urlstyle\endcsname\relax
  \providecommand{\doi}[1]{doi: #1}\else
  \providecommand{\doi}{doi: \begingroup \urlstyle{rm}\Url}\fi

\bibitem[Abay et~al.(2019)Abay, Zhou, Kantarcioglu, Thuraisingham, and Sweeney]{abay2019privacy}
Nazmiye~Ceren Abay, Yan Zhou, Murat Kantarcioglu, Bhavani Thuraisingham, and Latanya Sweeney.
\newblock {Privacy preserving synthetic data release using deep learning}.
\newblock In \emph{ECML PKDD}, 2019.

\bibitem[Acs et~al.(2018)Acs, Melis, Castelluccia, and De~Cristofaro]{acs2018differentially}
Gergely Acs, Luca Melis, Claude Castelluccia, and Emiliano De~Cristofaro.
\newblock {Differentially private mixture of generative neural networks}.
\newblock \emph{IEEE TKDE}, 2018.

\bibitem[{Alan Turing Institute}(2022)]{reprosyn2022}
{Alan Turing Institute}.
\newblock Reprosyn.
\newblock \url{https://github.com/alan-turing-institute/reprosyn}, 2022.

\bibitem[{Amazon AWS}(2022)]{aws2022how}
{Amazon AWS}.
\newblock {How to Evaluate the Quality of the Synthetic Data – Measuring from the Perspective of Fidelity, Utility, and Privacy}, 2022.

\bibitem[Annamalai et~al.(2024{\natexlab{a}})Annamalai, Gadotti, and Rocher]{annamalai2024linear}
Meenatchi Sundaram Muthu~Selva Annamalai, Andrea Gadotti, and Luc Rocher.
\newblock {A linear reconstruction approach for attribute inference attacks against synthetic data}.
\newblock In \emph{USENIX Security}, 2024{\natexlab{a}}.

\bibitem[Annamalai et~al.(2024{\natexlab{b}})Annamalai, Ganev, and De~Cristofaro]{annamalai2024you}
Meenatchi Sundaram Muthu~Selva Annamalai, Georgi Ganev, and Emiliano De~Cristofaro.
\newblock {``What do you want from theory alone?'' experimenting with tight auditing of differentially private synthetic data generation}.
\newblock In \emph{USENIX Security}, 2024{\natexlab{b}}.

\bibitem[Becker and Kohavi(1996)]{misc_adult_2}
Barry Becker and Ronny Kohavi.
\newblock {Adult}.
\newblock UCI Machine Learning Repository, 1996.

\bibitem[Berka et~al.(2000)]{berka}
P.~Berka et~al.
\newblock {Guide to the financial data set}.
\newblock PKDD2000 discovery challenge, 2000.

\bibitem[Borisov et~al.(2023)Borisov, Sessler, Leemann, Pawelczyk, and Kasneci]{borisov2023language}
Vadim Borisov, Kathrin Sessler, Tobias Leemann, Martin Pawelczyk, and Gjergji Kasneci.
\newblock {Language Models are Realistic Tabular Data Generators}.
\newblock In \emph{ICLR}, 2023.

\bibitem[Carlini et~al.(2022)Carlini, Chien, Nasr, Song, Terzis, and Tramèr]{carlini2022membership}
Nicholas Carlini, Steve Chien, Milad Nasr, Shuang Song, Andreas Terzis, and Florian Tramèr.
\newblock {Membership Inference Attacks From First Principles}.
\newblock In \emph{IEEE S\&P}, 2022.

\bibitem[Carlini et~al.(2023)Carlini, Hayes, Nasr, Jagielski, Sehwag, Tramer, Balle, Ippolito, and Wallace]{carlini2023extracting}
Nicolas Carlini, Jamie Hayes, Milad Nasr, Matthew Jagielski, Vikash Sehwag, Florian Tramer, Borja Balle, Daphne Ippolito, and Eric Wallace.
\newblock {Extracting training data from diffusion models}.
\newblock In \emph{USENIX Security}, 2023.

\bibitem[Ceritli et~al.(2023)Ceritli, Ghosheh, Chauhan, Zhu, Creagh, and Clifton]{ceritli2023synthesizing}
Taha Ceritli, Ghadeer~O Ghosheh, Vinod~Kumar Chauhan, Tingting Zhu, Andrew~P Creagh, and David~A Clifton.
\newblock {Synthesizing Mixed-type Electronic Health Records using Diffusion Models}.
\newblock \emph{arXiv:2302.14679}, 2023.

\bibitem[Chen et~al.(2020)Chen, Yu, Zhang, and Fritz]{chen2020gan}
Dingfan Chen, Ning Yu, Yang Zhang, and Mario Fritz.
\newblock {GAN-Leaks: A taxonomy of membership inference attacks against generative models}.
\newblock In \emph{ACM CCS}, 2020.

\bibitem[D'Amico et~al.(2023)D'Amico, Dall’Olio, Sala, et~al.]{damico2023synthetic}
Saverio D'Amico, Daniele Dall’Olio, Claudia Sala, et~al.
\newblock {Synthetic Data Generation by Artificial Intelligence to Accelerate Research and Precision Medicine in Hematology}.
\newblock \emph{JCO Clinical Cancer Informatics}, 2023.

\bibitem[De~Cristofaro(2024)]{cristofaro2024synthetic}
Emiliano De~Cristofaro.
\newblock {Synthetic Data: Methods, Use Cases, and Risks}.
\newblock \emph{IEEE S\&P Magazine}, 2024.

\bibitem[Duan et~al.(2023)Duan, Kong, Wang, Shi, and Xu]{duan2023diffusion}
Jinhao Duan, Fei Kong, Shiqi Wang, Xiaoshuang Shi, and Kaidi Xu.
\newblock {Are diffusion models vulnerable to membership inference attacks?}
\newblock In \emph{ICLR}, 2023.

\bibitem[Ganev and De~Cristofaro(2025)]{ganev2025inadequacysimilaritybasedprivacymetrics}
Georgi Ganev and Emiliano De~Cristofaro.
\newblock {The Inadequacy of Similarity-based Privacy Metrics: Privacy Attacks against ``Truly Anonymous'' Synthetic Datasets}.
\newblock In \emph{IEEE S\&P}, 2025.

\bibitem[Ganev et~al.(2025)Ganev, Annamalai, and De~Cristofaro]{ganev2025elusive}
Georgi Ganev, Meenatchi Sundaram Muthu~Selva Annamalai, and Emiliano De~Cristofaro.
\newblock {The Elusive Pursuit of Reproducing PATE-GAN: Benchmarking, Auditing, Debugging}.
\newblock \emph{TMLR}, 2025.

\bibitem[Golob et~al.(2025)Golob, Pentyala, Maratkhan, and De~Cock]{golob2025privacy}
Steven Golob, Sikha Pentyala, Anuar Maratkhan, and Martine De~Cock.
\newblock {Privacy Vulnerabilities in Marginals-based Synthetic Data}.
\newblock In \emph{SaTML}, 2025.

\bibitem[Gu{\'e}pin et~al.(2023)Gu{\'e}pin, Meeus, Cretu, and de~Montjoye]{guepin2023synthetic}
Florent Gu{\'e}pin, Matthieu Meeus, Ana-Maria Cretu, and Yves-Alexandre de~Montjoye.
\newblock {Synthetic is all you need: Removing the auxiliary data assumption for membership inference attacks against synthetic data}.
\newblock In \emph{ESORICS}, 2023.

\bibitem[Gu{\'e}pin et~al.(2024)Gu{\'e}pin, Kr{\v{c}}o, Meeus, and de~Montjoye]{guepin2024lost}
Florent Gu{\'e}pin, Nata{\v{s}}a Kr{\v{c}}o, Matthieu Meeus, and Yves-Alexandre de~Montjoye.
\newblock {Lost in the averages: A new specific setup to evaluate membership inference attacks against machine learning models}.
\newblock \emph{arXiv:2405.15423}, 2024.

\bibitem[Guillaudeux et~al.(2023)Guillaudeux, Rousseau, Petot, Bennis, Dein, Goronflot, Vince, Limou, Karakachoff, Wargny, and Gourraud]{guillaudeux2023patient}
Morgan Guillaudeux, Olivia Rousseau, Julien Petot, Zineb Bennis, Charles-Axel Dein, Thomas Goronflot, Nicolas Vince, Sophie Limou, Matilde Karakachoff, Matthieu Wargny, and Pierre-Antoine Gourraud.
\newblock {Patient-centric synthetic data generation, no reason to risk re-identification in biomedical data analysis}.
\newblock \emph{NPJ Digital Medicine}, 2023.

\bibitem[Hayes et~al.(2019)Hayes, Melis, Danezis, and De~Cristofaro]{hayes2019logan}
Jamie Hayes, Luca Melis, George Danezis, and Emiliano De~Cristofaro.
\newblock {LOGAN: membership inference attacks against generative models}.
\newblock In \emph{PoPETs}, 2019.

\bibitem[Hilprecht et~al.(2019)Hilprecht, H{\"a}rterich, and Bernau]{hilprecht2019monte}
Benjamin Hilprecht, Martin H{\"a}rterich, and Daniel Bernau.
\newblock {Monte Carlo and reconstruction membership inference attacks against generative models}.
\newblock In \emph{PoPETs}, 2019.

\bibitem[Homer et~al.(2008)Homer, Szelinger, Redman, Duggan, Tembe, Muehling, Pearson, Stephan, Nelson, and Craig]{homer2008mia}
Nils Homer, Szabolcs Szelinger, Margot Redman, David Duggan, Waibhav Tembe, Jill Muehling, John~V. Pearson, Dietrich~A. Stephan, Stanley~F. Nelson, and David~W. Craig.
\newblock {Resolving individuals contributing trace amounts of DNA to highly complex mixtures using high-density SNP genotyping microarrays}.
\newblock \emph{PLoS Genet}, 2008.

\bibitem[Houssiau et~al.(2022)Houssiau, Jordon, Cohen, Daniel, Elliott, Geddes, Mole, Rangel-Smith, and Szpruch]{houssiau2022tapas}
Florimond Houssiau, James Jordon, Samuel~N Cohen, Owen Daniel, Andrew Elliott, James Geddes, Callum Mole, Camila Rangel-Smith, and Lukasz Szpruch.
\newblock {Tapas: a toolbox for adversarial privacy auditing of synthetic data}.
\newblock \emph{NeurIPS Workshop on Synthetic Data for Empowering ML Research}, 2022.

\bibitem[Hu et~al.(2024)Hu, Wu, Li, Long, Garrido, Ge, Ding, Forsyth, Li, and Song]{hu2024sok}
Yuzheng Hu, Fan Wu, Qinbin Li, Yunhui Long, Gonzalo~Munilla Garrido, Chang Ge, Bolin Ding, David Forsyth, Bo~Li, and Dawn Song.
\newblock {Sok: Privacy-preserving data synthesis}.
\newblock In \emph{IEEE S\&P}, 2024.

\bibitem[Jordon et~al.(2018)Jordon, Yoon, and Van Der~Schaar]{jordon2018pate}
James Jordon, Jinsung Yoon, and Mihaela Van Der~Schaar.
\newblock {PATE-GAN: Generating synthetic data with differential privacy guarantees}.
\newblock In \emph{ICLR}, 2018.

\bibitem[Jordon et~al.(2022)Jordon, Szpruch, Houssiau, Bottarelli, Cherubin, Maple, Cohen, and Weller]{jordon2022syntheticdatawhat}
James Jordon, Lukasz Szpruch, Florimond Houssiau, Mirko Bottarelli, Giovanni Cherubin, Carsten Maple, Samuel~N Cohen, and Adrian Weller.
\newblock {Synthetic Data--what, why and how?}
\newblock \emph{arXiv:2205.03257}, 2022.

\bibitem[Kaabachi et~al.(2025)Kaabachi, Despraz, Meurers, Otte, Halilovic, Kulynych, Prasser, and Raisaro]{kaabachi2025scoping}
Bayrem Kaabachi, J{\'e}r{\'e}mie Despraz, Thierry Meurers, Karen Otte, Mehmed Halilovic, Bogdan Kulynych, Fabian Prasser, and Jean~Louis Raisaro.
\newblock {A scoping review of privacy and utility metrics in medical synthetic data}.
\newblock \emph{NPJ Digital Medicine}, 2025.

\bibitem[Kong et~al.(2023)Kong, Duan, Ma, Shen, Zhu, Shi, and Xu]{kong2023efficient}
Fei Kong, Jinhao Duan, RuiPeng Ma, Hengtao Shen, Xiaofeng Zhu, Xiaoshuang Shi, and Kaidi Xu.
\newblock {An efficient membership inference attack for the diffusion model by proximal initialization}.
\newblock \emph{arXiv:2305.18355}, 2023.

\bibitem[Kotelnikov et~al.(2023)Kotelnikov, Baranchuk, Rubachev, and Babenko]{kotelnikov2023tabddpm}
Akim Kotelnikov, Dmitry Baranchuk, Ivan Rubachev, and Artem Babenko.
\newblock {TabDDPM: Modelling Tabular Data with Diffusion Models}.
\newblock In \emph{ICML}, 2023.

\bibitem[Liu et~al.(2021)Liu, Vietri, and Wu]{liu2021iterative}
Terrance Liu, Giuseppe Vietri, and Steven~Z Wu.
\newblock {Iterative methods for private synthetic data: Unifying framework and new methods}.
\newblock \emph{NeurIPS}, 2021.

\bibitem[Liu et~al.(2023)Liu, Fan, Li, Tang, and Du]{liu2023tabular}
Tongyu Liu, Ju~Fan, Guoliang Li, Nan Tang, and Xiaoyong Du.
\newblock {Tabular Data Synthesis with Generative Adversarial Networks: Design Space and Optimizations}.
\newblock \emph{VLDBJ}, 2023.

\bibitem[Lu et~al.(2019)Lu, Wang, and Yu]{lu2019empirical}
Pei-Hsuan Lu, Pang-Chieh Wang, and Chia-Mu Yu.
\newblock {Empirical Evaluation on Synthetic Data Generation with Generative Adversarial Network}.
\newblock In \emph{WIMS}, 2019.

\bibitem[McKenna et~al.(2021)McKenna, Miklau, and Sheldon]{mckenna2021winning}
Ryan McKenna, Gerome Miklau, and Daniel Sheldon.
\newblock {Winning the NIST Contest: a scalable and general approach to differentially private synthetic data}.
\newblock \emph{JPC}, 2021.

\bibitem[McKenna et~al.(2022)McKenna, Mullins, Sheldon, and Miklau]{mckenna2022aim}
Ryan McKenna, Brett Mullins, Daniel Sheldon, and Gerome Miklau.
\newblock {Aim: An adaptive and iterative mechanism for differentially private synthetic data}.
\newblock \emph{PVLDB}, 2022.

\bibitem[Meeus et~al.(2023)Meeus, Guepin, Cre{\c{t}}u, and de~Montjoye]{meeus2023achilles}
Matthieu Meeus, Florent Guepin, Ana-Maria Cre{\c{t}}u, and Yves-Alexandre de~Montjoye.
\newblock {Achilles’ heels: vulnerable record identification in synthetic data publishing}.
\newblock In \emph{ESORICS}, 2023.

\bibitem[Mendelevitch and Lesh(2021)]{syntegra2021fidelity}
Ofer Mendelevitch and Michael~D Lesh.
\newblock {Fidelity and privacy of synthetic medical data}.
\newblock \emph{arXiv:2101.08658}, 2021.

\bibitem[Moro et~al.(2014)Moro, Rita, and Cortez]{bank_marketing_222}
S.~Moro, P.~Rita, and P.~Cortez.
\newblock {Bank Marketing}.
\newblock UCI Machine Learning Repository, 2014.

\bibitem[{Mostly AI}(2020)]{mostly2020truly}
{Mostly AI}.
\newblock {Truly anonymous synthetic data -- evolving legal definitions and technologies (part II)}.
\newblock \url{https://mostly.ai/blog/truly-anonymous-synthetic-data-legal-definitions-part-ii/}, 2020.

\bibitem[Mueller et~al.(2025)Mueller, Gruber, and Fok]{mueller2025cdtd}
Markus Mueller, Kathrin Gruber, and Dennis Fok.
\newblock {Continuous Diffusion for Mixed-Type Tabular Data}.
\newblock In \emph{ICLR}, 2025.

\bibitem[{Office for National Statistics}(2011)]{census2011}
{Office for National Statistics}.
\newblock Census microdata teaching files, 2011.

\bibitem[Panfilo(2022)]{panfilo2022generating}
Daniele Panfilo.
\newblock \emph{{Generating Privacy-compliant, Utility-preserving Synthetic Tabular and Relational Datasets through Deep Learning}}.
\newblock University of Trieste, 2022.

\bibitem[Panfilo et~al.(2023)Panfilo, Boudewijn, Saccani, Coser, Svara, Chauvenet, Mami, and Medvet]{panfilo2023a}
Daniele Panfilo, Alexander Boudewijn, Sebastiano Saccani, Andrea Coser, Borut Svara, Carlo Chauvenet, Ciro Mami, and Eric Medvet.
\newblock {A Deep Learning-based Pipeline for the Generation of Synthetic Tabular Data}.
\newblock \emph{IEEE Access}, 2023.

\bibitem[Pang et~al.(2024)Pang, Shafieinejad, Liu, Hazlewood, and He]{pang2024clavaddpm}
Wei Pang, Masoumeh Shafieinejad, Lucy Liu, Stephanie Hazlewood, and Xi~He.
\newblock {Clavaddpm: Multi-relational data synthesis with cluster-guided diffusion models}.
\newblock In \emph{NeurIPS}, 2024.

\bibitem[Park et~al.(2018)Park, Mohammadi, Gorde, Jajodia, Park, and Kim]{park2018data}
Noseong Park, Mahmoud Mohammadi, Kshitij Gorde, Sushil Jajodia, Hongkyu Park, and Youngmin Kim.
\newblock {Data Synthesis Based on Generative Adversarial Networks}.
\newblock \emph{PVLDB}, 2018.

\bibitem[Pastaltzidis et~al.(2022)Pastaltzidis, Dimitriou, Quezada-Tavarez, Aidinlis, Marquenie, Gurzawska, and Tzovaras]{pastaltzidis2022data}
Ioannis Pastaltzidis, Nikolaos Dimitriou, Katherine Quezada-Tavarez, Stergios Aidinlis, Thomas Marquenie, Agata Gurzawska, and Dimitrios Tzovaras.
\newblock {Data augmentation for fairness-aware machine learning: Preventing algorithmic bias in law enforcement systems}.
\newblock In \emph{ACM FAccT}, 2022.

\bibitem[Ping et~al.(2017)Ping, Stoyanovich, and Howe]{ping2017indhist}
Haoyue Ping, Julia Stoyanovich, and Bill Howe.
\newblock {DataSynthesizer: Privacy-Preserving Synthetic Datasets}.
\newblock In \emph{SSDBM}, 2017.

\bibitem[Pollock et~al.(2024)Pollock, Shilov, Dodd, and de~Montjoye]{pollock2024free}
Joseph Pollock, Igor Shilov, Euodia Dodd, and Yves-Alexandre de~Montjoye.
\newblock {Free Record-Level Privacy Risk Evaluation Through Artifact-Based Methods}.
\newblock \emph{arXiv:2411.05743}, 2024.

\bibitem[Pyrgelis et~al.(2017)Pyrgelis, Troncoso, and De~Cristofaro]{pyrgelis2017knockknockwhosthere}
Apostolos Pyrgelis, Carmela Troncoso, and Emiliano De~Cristofaro.
\newblock {Knock Knock, Who's There? Membership Inference on Aggregate Location Data}.
\newblock In \emph{NDSS}, 2017.

\bibitem[Salem et~al.(2023)Salem, Cherubin, Evans, K{\"o}pf, Paverd, Suri, Tople, and Zanella-B{\'e}guelin]{salem2023sok}
Ahmed Salem, Giovanni Cherubin, David Evans, Boris K{\"o}pf, Andrew Paverd, Anshuman Suri, Shruti Tople, and Santiago Zanella-B{\'e}guelin.
\newblock {SoK: Let the privacy games begin! A unified treatment of data inference privacy in machine learning}.
\newblock In \emph{IEEE S\&P}, 2023.

\bibitem[Shi et~al.(2025)Shi, Xu, Hua, Zhang, Ermon, and Leskovec]{shi2025tabdiff}
Juntong Shi, Minkai Xu, Harper Hua, Hengrui Zhang, Stefano Ermon, and Jure Leskovec.
\newblock {TabDiff: a Mixed-type Diffusion Model for Tabular Data Generation}.
\newblock In \emph{ICLR}, 2025.

\bibitem[Shokri et~al.(2017)Shokri, Stronati, Song, and Shmatikov]{shokri2017membership}
Reza Shokri, Marco Stronati, Congzheng Song, and Vitaly Shmatikov.
\newblock {Membership inference attacks against machine learning models}.
\newblock In \emph{IEEE S\&P}, 2017.

\bibitem[Sivakumar et~al.(2023)Sivakumar, Ramamurthy, Radhakrishnan, and Won]{sivakumar2023generativemtd}
Jayanth Sivakumar, Karthik Ramamurthy, Menaka Radhakrishnan, and Daehan Won.
\newblock {GenerativeMTD: A Deep Synthetic Data Generation Framework for Small Datasets}.
\newblock \emph{KBS}, 2023.

\bibitem[Stadler et~al.(2022)Stadler, Oprisanu, and Troncoso]{stadler2022synthetic}
Theresa Stadler, Bristena Oprisanu, and Carmela Troncoso.
\newblock {Synthetic data -- anonymization groundhog day}.
\newblock In \emph{USENIX Security}, 2022.

\bibitem[{Synthetic Data Expert Group, FCA}(2024)]{fca_synth}
{Synthetic Data Expert Group, FCA}.
\newblock Using synthetic data in financial services.
\newblock \url{https://www.fca.org.uk/publication/corporate/report-using-synthetic-data-in-financial-services.pdf}, 2024.

\bibitem[{Syntho}(2025)]{syntho2025report}
{Syntho}.
\newblock \url{https://www.syntho.ai/synthos-quality-assurance-report/}, 2025.

\bibitem[van Breugel et~al.(2023)van Breugel, Sun, Qian, and van~der Schaar]{van2023membership}
Boris van Breugel, Hao Sun, Zhaozhi Qian, and Mihaela van~der Schaar.
\newblock {Membership inference attacks against synthetic data through overfitting detection}.
\newblock \emph{AISTATS}, 2023.

\bibitem[{Vector Institute}(2025)]{midst}
{Vector Institute}.
\newblock Midstmodels.
\newblock \url{https://github.com/VectorInstitute/MIDST/}, 2025.

\bibitem[Venugopal et~al.(2022)Venugopal, Shafqat, Venugopal, Tillbury, Stafford, and Bourazeri]{venugopal2022privacy}
Rohit Venugopal, Noman Shafqat, Ishwar Venugopal, Benjamin Mark~John Tillbury, Harry~Demetrios Stafford, and Aikaterini Bourazeri.
\newblock {Privacy Preserving Generative Adversarial Networks to Model Electronic Health Records}.
\newblock \emph{Neural Networks}, 2022.

\bibitem[Wu et~al.(2025)Wu, Pang, Liu, and Wu]{wu2025winning}
Xiaoyu Wu, Yifei Pang, Terrance Liu, and Steven Wu.
\newblock {Winning the MIDST Challenge: New Membership Inference Attacks on Diffusion Models for Tabular Data Synthesis}.
\newblock \emph{arXiv:2503.12008}, 2025.

\bibitem[Xie et~al.(2018)Xie, Lin, Wang, Wang, and Zhou]{xie2018differentially}
Liyang Xie, Kaixiang Lin, Shu Wang, Fei Wang, and Jiayu Zhou.
\newblock {Differentially private generative adversarial network}.
\newblock \emph{arXiv:1802.06739}, 2018.

\bibitem[Xu et~al.(2019)Xu, Skoularidou, Cuesta-Infante, and Veeramachaneni]{ctgan}
Lei Xu, Maria Skoularidou, Alfredo Cuesta-Infante, and Kalyan Veeramachaneni.
\newblock Modeling tabular data using conditional gan.
\newblock In \emph{NeurIPS}, 2019.

\bibitem[Yale et~al.(2019)Yale, Dash, Dutta, Guyon, Pavao, and Bennett]{yale2019assessing}
Andrew Yale, Saloni Dash, Ritik Dutta, Isabelle Guyon, Adrien Pavao, and Kristin~P Bennett.
\newblock {Assessing Privacy and Quality of Synthetic Health Data}.
\newblock In \emph{AIDR}, 2019.

\bibitem[{YData}(2023)]{ydata2023how}
{YData}.
\newblock {How to evaluate the re-identification risk in Synthetic Data?}, 2023.

\bibitem[Ye et~al.(2022)Ye, Maddi, Murakonda, Bindschaedler, and Shokri]{ye2022enhanced}
Jiayuan Ye, Aadyaa Maddi, Sasi~Kumar Murakonda, Vincent Bindschaedler, and Reza Shokri.
\newblock {Enhanced membership inference attacks against machine learning models}.
\newblock In \emph{ACM CCS}, 2022.

\bibitem[Yoon et~al.(2023)Yoon, Mizrahi, Ghalaty, and others.]{yoon2023ehr}
Jinsung Yoon, Michel Mizrahi, Nahid~Farhady Ghalaty, and others.
\newblock {EHR-Safe: Generating High-fidelity and Privacy-preserving Synthetic Electronic Health Records}.
\newblock \emph{NPJ Digital Medicine}, 2023.

\bibitem[Zarifzadeh et~al.(2024)Zarifzadeh, Liu, and Shokri]{zarifzadehlow}
Sajjad Zarifzadeh, Philippe Liu, and Reza Shokri.
\newblock {Low-Cost High-Power Membership Inference Attacks}.
\newblock In \emph{ICML}, 2024.

\bibitem[Zhang et~al.(2024)Zhang, Zhang, Shen, Srinivasan, Qin, Faloutsos, Rangwala, and Karypis]{zhang2024tabsyn}
Hengrui Zhang, Jiani Zhang, Zhengyuan Shen, Balasubramaniam Srinivasan, Xiao Qin, Christos Faloutsos, Huzefa Rangwala, and George Karypis.
\newblock {Mixed-Type Tabular Data Synthesis with Score-based Diffusion in Latent Space}.
\newblock In \emph{ICLR}, 2024.

\bibitem[Zhang et~al.(2017)Zhang, Cormode, Procopiuc, Srivastava, and Xiao]{baynet}
Jun Zhang, Graham Cormode, Cecilia~M Procopiuc, Divesh Srivastava, and Xiaokui Xiao.
\newblock {Privbayes: Private data release via bayesian networks}.
\newblock \emph{ACM TODS}, 2017.

\bibitem[Zhang et~al.(2023)Zhang, Wang, Yan, Li, and Liu]{zhang2023generative}
Tianping Zhang, Shaowen Wang, Shuicheng Yan, Jian Li, and Qian Liu.
\newblock {Generative Table Pre-training Empowers Models for Tabular Prediction}.
\newblock In \emph{EMNLP}, 2023.

\bibitem[Zhang et~al.(2018)Zhang, Ji, and Wang]{zhang2018differentially}
Xinyang Zhang, Shouling Ji, and Ting Wang.
\newblock {Differentially private releasing via deep generative model (technical report)}.
\newblock \emph{arXiv:1801.01594}, 2018.

\bibitem[Zhao et~al.(2021)Zhao, Kunar, Birke, and Chen]{zhao2021ctab}
Zilong Zhao, Aditya Kunar, Robert Birke, and Lydia~Y Chen.
\newblock {CTAB-GAN: Effective Table Data Synthesizing}.
\newblock In \emph{ACML}, 2021.

\bibitem[Zhu et~al.(2023)Zhu, Chen, Grossklags, and Fritz]{zhu2023data}
Derui Zhu, Dingfan Chen, Jens Grossklags, and Mario Fritz.
\newblock {Data forensics in diffusion models: A systematic analysis of membership privacy}.
\newblock \emph{arXiv:2302.07801}, 2023.

\end{thebibliography}
\bibliographystyle{plainnat}

\end{document}